\documentclass[a4paper,fleqn,10pt]{article}

\usepackage{bbold, epsfig}
\usepackage{latexsym, amsmath, euscript,enumitem}

\usepackage[T1]{fontenc}
\usepackage{fancyhdr}

\newcommand{\esssup}[1]{\mathop{\rm ess\ sup}}
\newcommand{\essinf}[1]{\mathop{\rm ess\ inf}}
\newcommand{\N}{{\rm I\kern - 2.5pt N}}
\newcommand{\Z}{{\rm Z\kern - 5.5pt Z}}
\newcommand{\Q}{{\rm I\kern - 5.25pt Q}}
\newcommand{\C}{{\rm I\kern - 6.25pt C}}
\newcommand{\R}{{\rm I\kern - 2.5pt R}}

\newcommand{\bbf}{\mathbf{b}}
\newcommand{\dbf}{\mathbf{d}}
\newcommand{\Dbf}{\mathbf{D}}
\newcommand{\gbf}{\mathbf{g}}
\newcommand{\Ibf}{\mathbf{I}}
\newcommand{\jbf}{\mathbf{j}}
\newcommand{\Jbf}{\mathbf{J}}
\newcommand{\nbf}{\mathbf{n}}
\newcommand{\Pbf}{\mathbf{P}}
\newcommand{\qbf}{\mathbf{q}}
\newcommand{\rbf}{\mathbf{r}}
\newcommand{\Sbf}{\mathbf{S}}
\newcommand{\ubf}{\mathbf{u}}
\newcommand{\vbf}{\mathbf{v}}
\newcommand{\xbf}{\mathbf{x}}
\newcommand{\Phibf}{\mathbf{\Phi}}
\newcommand{\na}{\nabla}
\newcommand{\pa}{\partial}
\begin{document}
\title{\bf On the multi-physics of mass-transfer\\ across fluid interfaces}
\author{Dieter Bothe\\
Center of Smart Interfaces\\
TU Darmstadt\\
bothe@csi.tu-darmstadt.de}
\date{}
\maketitle

\begin{abstract}
\noindent
Mass transfer of gaseous components from rising bubbles to the ambient liquid can be described based on continuum mechanical sharp-interface balances of mass, momentum and species mass. In this context, the standard model consists of the two-phase Navier-Stokes equations for incompressible fluids with constant surface tension, complemented by reaction-advection-diffusion equations for all constituents, employing Fick's law. This standard model is inconsistent with the continuity equation, the momentum balance and the second law of thermodynamics. The present paper reports on the details of these severe shortcomings and provides thermodynamically consistent model extensions which are required to capture various phenomena which occur due to the multi-physics of interfacial mass transfer. In particular, we provide a simple derivation of the interface Maxwell-Stefan equations which does not require a time scale separation, while the main contribution is to show how interface concentrations and interface chemical potentials mediate the influence on mass transfer of a transfer component exerted by the change in interface energy due to an adsorbing surfactant.
\end{abstract}


\section*{Introduction}
The standard model for the continuum mechanical description of mass transfer across fluid interfaces is based on the incompressible two-phase Navier-Stokes equations for fluid systems without phase change. To be more precise, inside the fluid phases the governing equations are
\begin{align}
	&\na \cdot \vbf=0,\label{E1}\\
	&\pa_t (\rho\vbf)+ \na \cdot (\rho \vbf \otimes \vbf)+\na p=\na \cdot \Sbf^{\rm visc}+\rho\gbf \label{E2}
\end{align}
with the viscous stress tensor
\begin{equation}\label{E3}
\Sbf^{\rm visc}=\eta(\na\vbf+\na \vbf^{\sf T}),
\end{equation}
where the material parameters depend on the respective phase. Whenever distinction between the different phases is necessary, $+$ and $-$ are used as phase indices. Since phase change is neglected in the standard model, there are no convective fluxes across the interface, i.e.\ the normal component $V_\Sigma=\vbf^\Sigma \cdot \nbf_\Sigma$
of the interfacial velocity coincides with the normal component of both the adjacent fluid velocities. Moreover, since both fluids have non-vanishing viscosity, no-slip between the two phases at the interface is usually imposed. Finally, constant surface tension is assumed. In this situation, the interfacial jump conditions for total mass and momentum are
\begin{equation}\label{E4}
[\![\vbf]\!]=0, \quad [\![-\Sbf]\!] \cdot \nbf_\Sigma=\sigma \kappa_\Sigma \nbf_\Sigma,
\end{equation}
where $\Sbf=-p \Ibf+\Sbf^{\rm visc}$ is the stress tensor and $\kappa_\Sigma=-\na \cdot \nbf_\Sigma$ is twice the mean curvature of the interface. Here $\nbf_\Sigma$ is the unit normal at the interface directed into the bulk phase --, say, and the notation
\begin{equation}\label{E5}
[\![\phi]\!](\xbf)=\lim\limits_{h \to 0+} \big(\phi(\xbf+h\nbf_\Sigma)-\phi(\xbf-h\nbf_\Sigma)\big)
\end{equation}
stands for the jump of a field $\phi$ across the interface.

The local molar concentration $c_i$ of a chemical species $A_i$ is governed by the balance equation
\begin{equation}\label{E6}
\pa_t c_i+\na \cdot (c_i \vbf+\Jbf_i)=r_i,
\end{equation}
where the molecular fluxes $\Jbf_i$ are typically modeled according to Fick's law as
\begin{equation}\label{E7}
\Jbf_i=-D_i \na c_i
\end{equation}
with constant diffusivity. The source term on the right-hand side in (\ref{E6}) accounts for chemical reactions. At the interface, the diffusive fluxes in normal direction are commonly supposed to be continuous, i.e.\ \begin{equation}\label{E8}
[\![-D_i \na c_i]\!] \cdot \nbf_\Sigma=0.
\end{equation}
One more constitutive equation is needed to determine the concentration profiles of $A_i$, where instantaneous local chemical equilibrium at the interface is usually employed. This means continuity of the chemical potentials at the interface, i.e.\ \begin{equation}\label{E9}
[\![\mu_i]\!] =0.
\end{equation}
Using standard relations for the $\mu_i$, this leads to Henry's law. In the form for molar concentrations, the latter states that
\begin{equation}\label{E10}
c^-_k=c^+_k/H_k
\end{equation}
with a Henry coefficient $H_k$, often assumed to be constant. Equations (\ref{E1})--(\ref{E4}) and (\ref{E6})--(\ref{E10}) comprise what is called the ``standard model'' throughout this paper. This standard model, sometimes with further simplifications like homogeneous gas phase concentrations or even constant liquid-sided concentration at the interface, is the basis of almost all detailed numerical simulations of mass transfer across fluid interfaces up to now; see \cite{1}, \cite{2} and the extensive list of references given there. For fundamentals on continuum mechanical modeling of two-phase fluid systems we refer in particular to \cite{3, 4, Bedeaux, 5}.

The remainder of the present paper reports on major short-comings of the standard model and explains how to overcome these in the framework of continuum thermodynamics. For this purpose, we start on the interface, proceed to the adjacent bulk layers before we visit the bulk phases and, finally, the outer boundary of the domain in which the balance equations are studied. To enable a more descriptive notion, we focus on mass transfer from a bubble to the ambient liquid as a prototype case, but all considerations are also valid for other phase topologies (drops, films etc.) and for liquid-liquid systems.
\section*{The Fluid Interface}
In our continuum thermodynamical description of the interface, a sharp-interface model is employed which represents a thin layer in which the partial mass densities change from one to the other (local) bulk value within a layer of thickness $\delta$ in the order of a few \AA ngström. Since every transfer component $A_i$ of the mixture, in principle every constituent, is also present in this layer, it has a non-zero interface concentration $c^\Sigma_i$, hence also an interface chemical potential $\mu^\Sigma_i$ in the sharp-interface model. The mass balance for $A_i$ on the interface then becomes more involved and reads
\begin{equation}\label{E11}
\pa^\Sigma_t c^\Sigma_i + \na_\Sigma \cdot (c^\Sigma_i \vbf^\Sigma + \Jbf^\Sigma_i)+ [\![c_i(\vbf-\vbf^\Sigma)+\Jbf_i]\!] \cdot \nbf_\Sigma=\rbf^\Sigma_i,
\end{equation}
where $\vbf^\Sigma$ denotes the interfaces barycentric velocity, $\Jbf^\Sigma_i$ the interfacial diffusive flux and $r^\Sigma_i$ is the total molar rate of change of $A_i$ due to interface chemical reactions between the species. Furthermore, $\pa^\Sigma_t c^\Sigma_i$ denotes the time derivative -- here of interface concentrations -- along a path which follows the interface's normal motion; note that a partial time derivative of interface quantities does not exist in the usual sense if the interface moves. Observe that the surface divergence in (\ref{E11}) contains a curvature dependent contribution since
\begin{equation}\label{E12}
\na_\Sigma \cdot \vbf^\Sigma= \na_\Sigma \cdot \vbf^\Sigma_{||}-V_\Sigma \kappa_\Sigma \nbf_\Sigma,
\end{equation}
where the subscript $||$ denotes the tangential part. Note also that even for vanishing interface concentrations and without interface chemistry, equation (\ref{E11}) does not reduce to (\ref{E8}) from the standard model, but the resulting jump condition also contains a convective part due to a relative motion of the interface to the bulk matter. This is crucial, for example, to model the condensation of a vapor bubble or the dissolution of a pure gas bubble.

If (\ref{E11}) is multiplied by the molar mass $M_i$, then the partial mass balance
\begin{equation}\label{E13}
\pa^\Sigma_t \rho^\Sigma_i + \na_\Sigma \cdot (\rho^\Sigma_i \vbf^\Sigma + \jbf^\Sigma_i)+ [\![\rho_i(\vbf-\vbf^\Sigma)+\jbf_i]\!]\cdot \nbf_\Sigma= M_i r^\Sigma_i
\end{equation}
results with $\jbf^\Sigma_i=M_i \Jbf^\Sigma_i$ the diffusive mass flux. It is helpful to write out the jump-bracket, i.e.\ \begin{equation}\label{E14}
[\![\rho_i(\vbf-\vbf^\Sigma)+\jbf_i]\!] \cdot \nbf_\Sigma=-(\rho^+_i(\vbf^+-\vbf^\Sigma)+\jbf^+_i) \cdot \nbf^+-(\rho^-_i(\vbf^- -\vbf^\Sigma)+ \jbf^-_i) \cdot \nbf^-
\end{equation}
with $\nbf^\pm$ denoting the outer unit normals to the respective bulk phases. We abbreviate the terms on the right-hand side by letting
\begin{equation}\label{E15}
\dot{m}^{\pm,\Sigma}_i= (\rho^\pm_i(\vbf^\pm-\vbf^\Sigma)+ \jbf^\pm_i) \cdot \nbf^\pm,
\end{equation}
since these terms denote the mass transfer rates from the respective bulk phase to the interface and comprise the central object of the present paper. Guided by the case of a (soluble) surfactant, i.e.\ a chemical species which significantly accumulates at the interface, thereby changing the interface energy, the one-sided mass transfer terms are further split into an adsorption and a desorption term according to
\begin{equation}\label{E16}
\dot{m}^{+,\Sigma}_i=s^{ad,+}_i -s^{de,+}_i, \quad \dot{m}^{-,\Sigma}_i= s^{ad,-}_i -s^{de,-}_i.
\end{equation}
The interfacial balance for total mass reads as
\begin{equation}\label{E17}
\pa^\Sigma_t \rho^\Sigma+ \na_\Sigma \cdot(\rho^\Sigma \vbf^\Sigma)+[\![\rho(\vbf-\vbf^\Sigma)]\!] \cdot \nbf_\Sigma=0,
\end{equation}
and we abbreviate the total mass transfer rate from the respective bulk phase to the interface as $\dot{m}^{\pm,\Sigma}=\rho^\pm(\vbf^\pm-\vbf^\Sigma)\cdot \nbf^\pm$,
while inside the jump bracket $\dot{m}$ stands for $\dot{m}=\rho (\vbf-\vbf^\Sigma) \cdot \nbf_\Sigma$ with corresponding one-sided limits, but with $\nbf_\Sigma$ instead of $\nbf^\pm$.
Summation of all partial mass balances for the interface hence shows that
\begin{equation}\label{E18}
\sum\limits^N_{i=1} \jbf^\Sigma_i=0
\end{equation}
always holds for the diffusive mass fluxes. The latter implies a corresponding constraint for the diffusive molar mass fluxes which is not satisfied by Fickean fluxes according to (\ref{E7}).

Equation (\ref{E17}) replaces the normal component of the first equation in (\ref{E4}), while the general form of the second equation in (\ref{E4}) is the interfacial momentum balance
\begin{equation}\label{E19}
\pa^\Sigma_t(\rho^\Sigma \vbf^\Sigma)+ \na_\Sigma \cdot(\rho^\Sigma \vbf^\Sigma \otimes \vbf^\Sigma)+ [\![\rho \vbf \otimes (\vbf-\vbf^\Sigma)-\Sbf]\!] \cdot \nbf_\Sigma=\na_\Sigma \cdot \Sbf^\Sigma + \rho^\Sigma {\bf b}^\Sigma
\end{equation}
with specific body force ${\bf b}^\Sigma$ and the surface stress tensor $\Sbf^\Sigma$ which contains the surface tension $\sigma$, commonly used instead of the surface pressure
$p^\Sigma = -\sigma$, and viscous interface stresses due to intrinsic surface viscosities.
For a fluid interface with Newtonian rheology this leads to the Boussinesq-Scriven surface stress, i.e.\ \begin{equation}\label{E20}
\Sbf^\Sigma=(\sigma+(\eta^\Sigma_d-\eta^\Sigma_s) \na_\Sigma \cdot \vbf^\Sigma) \Pbf_\Sigma+ 2\eta^\Sigma_s \Dbf^\Sigma
\end{equation}
with the interface dilatational and shear viscosities $\eta^\Sigma_d \geq \eta^\Sigma_s \geq 0$, the surface projector
\begin{equation}\label{E21}
\Pbf_\Sigma=\Ibf-\nbf_\Sigma \otimes \nbf_\Sigma
\end{equation}
and the surface rate-of-deformation tensor $\Dbf^\Sigma= \frac{1}{2}\Pbf_\Sigma(\na_\Sigma \vbf^\Sigma+(\na_\Sigma \vbf^\Sigma)^{\sf T})\Pbf_\Sigma$. For further information about interfacial rheology we refer to \cite{4}, while in the present work we rest content with a reduced form which assumes that, while the interface contributions in the partial \emph{mass} balances are of crucial importance, their \emph{inertia} can be neglected and that no intrinsic interface viscosities appear. Then the interfacial momentum balance becomes a momentum transmission condition which reads as
\begin{equation}\label{E22}
[\![\dot{m}\vbf]\!]=[\![\Sbf]\!] \cdot \nbf_\Sigma+ \sigma\kappa_\Sigma \nbf_\Sigma+\na_\Sigma \sigma.
\end{equation}
Note that (\ref{E22}) is an interfacial force balance which accounts for the effects from mass transfer (the Stefan flow on the left-hand side) and for Marangoni effects expressed by the last term on the right-hand side. If the fluids are stagnant, i.e.\ if no flow occurs, then (\ref{E22}) reduces to
\begin{equation}\label{E23}
[\![p]\!] \nbf_\Sigma=\sigma\kappa_\Sigma \nbf_\Sigma+ \na_\Sigma \sigma.
\end{equation}
The normal component of (\ref{E23}) implies the Young-Laplace law, i.e.\ \begin{equation}\label{E24}
p^+-p^-= \frac{2\sigma}{R}
\end{equation}
in case of spherical bubbles or droplets of radius $R$, while the tangential part implies constant surface tension (individually on connected interface components), since there is no term to balance the tangentially acting Marangoni stress. The important consequence of this is
\begin{equation}\label{E25}
\na_\Sigma \sigma \not=0 \quad\Rightarrow\quad \vbf \not=0,
\end{equation}
i.e.\ any Marangoni stress induces a flow at the interface, the so-called Marangoni convection, which usually has a significant impact on mass transfer processes.

To obtain a closed model, i.e.\ a system of partial differential equations containing rates and fluxes which are expressed in terms of the primitive (balanced) quantities up to some material parameters which must be determined from measurements or a micro-theory, one needs constitutive equations for the unknown rates and fluxes. At the interface, these include the individual one-sided mass fluxes $\dot{m}^{\pm,\Sigma}_i$. Moreover, one needs constitutive equations for the interface chemical potentials and an interface equation of state modeling the surface tension. The latter quantities all follow from a constitutive equation for the interface free energy $\rho^\Sigma \psi^\Sigma$ via the surface Gibbs-Duhem equation
\begin{equation}\label{E26}
\rho^\Sigma \psi^\Sigma = \sigma +  \sum\limits^N_{i=1} \rho^\Sigma_i \mu^\Sigma_i
\end{equation}
and the fundamental differential relation
\begin{equation}\label{E27}
d(\rho^\Sigma \psi^\Sigma)=-\rho^\Sigma s^\Sigma dT^\Sigma+ \sum\limits_{i=1}^N \mu^\Sigma_i d\rho^\Sigma_i,
\end{equation}
where $s^\Sigma$ denotes the specific interfacial entropy. The latter implies in particular
\begin{equation}\label{E28}
\mu^\Sigma_i= \frac{\pa(\rho^\Sigma \psi^\Sigma)}{\pa\rho^\Sigma_i} \quad\text{ for }\quad \rho^\Sigma \psi^\Sigma=\rho^\Sigma \psi^\Sigma(T^\Sigma, \rho^\Sigma_1, \dots, \rho^\Sigma_N).
\end{equation}
The constitutive equations should of course be consistent with the second law of thermodynamics, i.e.\ such that the entropy inequality is satisfied: $\zeta^\Sigma \geq 0$ for any thermodynamic process. Now, in the considered case with partial mass densities on the interface, the interfacial entropy production $\zeta^\Sigma$ is given by
\begin{align}
	\zeta^\Sigma &= \qbf^\Sigma \cdot \na_\Sigma \frac{1}{T^\Sigma}- \sum\limits^N_{i=1} \jbf^\Sigma_i \cdot \left(\na_\Sigma \frac{\mu^\Sigma_i}{T^\Sigma}- \frac{\bbf^\Sigma_i}{T^\Sigma}\right)- \sum\limits^{N_R}_{a=1} R^\Sigma_a \mathcal{A}^\Sigma_a \label{E29}\\
	&- \frac{1}{T^\Sigma}(\vbf^+-\vbf^\Sigma)_{||} \cdot (\Sbf^+ \cdot \nbf^+)_{||}- \frac{1}{T^\Sigma}(\vbf^- - \vbf^\Sigma)_{||} \cdot (\Sbf^- \cdot \nbf^-)_{||} \nonumber\\
	&+ \left(\frac{1}{T^\Sigma}- \frac{1}{T^+}\right) \left(\dot{m}(e^+ + \frac{p^+}{\rho^+})+ \qbf^+ \cdot \nbf^+\right) \nonumber\\
&+  \left(\frac{1}{T^\Sigma}- \frac{1}{T^-}\right) \left(\dot{m}(e^- + \frac{p^-}{\rho^-})+ \qbf^- \cdot \nbf^-\right) \nonumber\\
	&+\sum\limits^N_{i=1} \dot{m}^{+,\Sigma}_i \left(\frac{\mu^+_i}{T^+}- \frac{\mu^\Sigma_i}{T^\Sigma}+ \frac{1}{T^\Sigma} \left(\frac{(\vbf^+-\vbf^\Sigma)^2}{2}- \nbf^+ \cdot \frac{\Sbf^{+,\rm visc}}{\rho^+} \cdot \nbf^+\right)\right) \nonumber\\
	&+ \sum\limits^N_{i=1} \dot{m}^{-,\Sigma}_i \left(\frac{\mu^-_i}{T^-}- \frac{\mu^\Sigma_i}{T^\Sigma}+ \frac{1}{T^\Sigma} \left(\frac{(\vbf^- -\vbf^\Sigma)^2}{2}- \nbf^- \cdot \frac{\Sbf^{-,\rm visc}}{\rho^-} \cdot \nbf^-\right)\right). \nonumber
\end{align}
A derivation of (\ref{E29}) along the lines of classical TIP can be found in \cite{Bedeaux},
where the relation (\ref{E26}) is imposed. The interfacial entropy production has been derived in different form before
in \cite{Kovac}, based on the single component case treated in \cite{BAM}.
The Gibbs-Duhem relation (\ref{E26}) can also be obtained as a consequence and we briefly explain the main steps in an appendix. The interfacial balances for partial mass, total momentum and energy can also be found in \cite{5} and \cite{6}. While in \cite{6} the entropy production is only considered in the special case with $\rho^\Sigma_i=0$, the interfacial entropy production in \cite{5} is differently structured and the one-sided partial mass transfer rates $\dot{m}^{\pm,\Sigma}_i$ do not appear as separate quantities. A similar form of the interfacial entropy production can be found in \cite{7, 8}, again without terms containing the one-sided partial mass transfer rates $\dot{m}^{\pm,\Sigma}_i$. At this point it is also important to note that in \cite{4, 7, 8, 5} the concept of Gibbs excess quantities is employed, which is known to lead to problems like negative excess of interface partial mass. Employing this concept, a quantity which is not accumulated at the interface would typically receive no excess mass which implies that its mass transfer will not involve an interface chemical potential. But the latter is shown below to account for the influence of changes in the interfacial free energy.

Each binary product in (\ref{E29}) corresponds to a dissipative physical mechanism at the interface. In each product, one factor is a constitutive quantity which is to be modeled in dependence of its co-factor. For processes not too far from equilibrium, a linear (in the co-factors) closure is appropriate. In this way the first product, which relates to heat conduction, leads to Fourier's law for the interfacial heat flux, i.e.\ \begin{equation}\label{E30}
\qbf^\Sigma= -\lambda^\Sigma \na_\Sigma T^\Sigma \quad\text{ with }\quad \lambda^\Sigma \geq 0.
\end{equation}
The interface temperature $T^\Sigma$ is, in general, different from the one-sided limits of the bulk temperature. Note also that $\lambda^\Sigma$ is allowed to depend on $T^\Sigma$ and the $\rho^\Sigma_i$. The latter is also true for all further closure parameters which are introduced below without explicit mentioning.

The second product in (\ref{E29}) corresponds to multicomponent diffusion on the interface. The standard linear closure requires first to incorporate the constraint (\ref{E18}), say by eliminating the flux $\jbf^\Sigma_N$. This leads to the mass diffusion fluxes
\begin{equation}\label{E31}
\jbf^\Sigma_i=- \sum\limits^{N-1}_{k=1} L_{ik} \cdot \left(\na_\Sigma \frac{\mu^\Sigma_k- \mu^\Sigma_N}{T^\Sigma}- \frac{\bbf^\Sigma_k-\bbf^\Sigma_N}{T^\Sigma}\right),
\end{equation}
where the matrix $[L_{ik}]$ of mobilities is positive definite. It can also be shown that $[L_{ik}]$ is symmetric, i.e.\ the Onsager symmetries hold; see \cite{9} for a rigorous theory of Onsager symmetries for transport coefficients. A difficulty with this generalized Fickean form is the fact that the mobilities are complicated functions of the composition - this needs to be the case, since for constant mobilities the positivity of solutions would not be guaranteed. For this reason the Maxwell-Stefan form of surface diffusion is advantageous, and we provide a simple derivation based on (\ref{E26}), (\ref{E27}) and (\ref{E29}), being fully rigorous in the isothermal case. At this point it should be noted that the diffusion flux is related to a diffusion velocity $\ubf^\Sigma_i$ via
\begin{equation}\label{E32}
\jbf^\Sigma_i= \rho^\Sigma_i \ubf^\Sigma_i:= \rho^\Sigma_i(\vbf^\Sigma_i-\vbf^\Sigma),
\end{equation}
where $\vbf^\Sigma_i$ is the continuum mechanical velocity of the individual component $A_i$. Then the starting point is the following reformulation of the relevant entropy production contribution. We have
\begin{equation}\label{E33}
\zeta^\Sigma_{\rm DIFF}=-\sum\limits^N_{i=1} \ubf^\Sigma_i \cdot \left(\rho^\Sigma_i \na_\Sigma \frac{\mu^\Sigma_i}{T^\Sigma}- \frac{\rho^\Sigma_i \bbf^\Sigma_i}{T^\Sigma}- \rho^\Sigma_i \Lambda\right)
\end{equation}
with a Lagrange parameter $\Lambda$ which can be arbitrary due to (\ref{E18}). This parameter is chosen in such a way that the co-factors in (\ref{E33}) sum up to zero, hence
\begin{equation}\label{E34}
\Lambda= \frac{1}{\rho^\Sigma} \sum\limits_{i=1}^N \left(\rho_i^\Sigma \na_\Sigma \frac{\mu^\Sigma_i}{T^\Sigma}- \frac{\rho^\Sigma_i \bbf^\Sigma_i}{T^\Sigma}\right).
\end{equation}
Using (\ref{E26}) and (\ref{E27}), a straightforward computation shows that (\ref{E33}) then becomes
\begin{equation}\label{E35}
\zeta^\Sigma_{\rm DIFF}=- \sum\limits^N_{i=1} \ubf^\Sigma_i \cdot \dbf^\Sigma_i
\end{equation}
with the generalized thermodynamic interface driving forces
\begin{equation}\label{E36}
\dbf^\Sigma_i=\rho^\Sigma_i \na_\Sigma \frac{\mu^\Sigma_i}{T^\Sigma}-\rho^\Sigma_i \frac{\bbf^\Sigma_i-\bbf^\Sigma}{T^\Sigma}+ \frac{y^\Sigma_i}{T^\Sigma} \na_\Sigma \sigma-y^\Sigma_i h^\Sigma \na_\Sigma \frac{1}{T^\Sigma},
\end{equation}
where $y^\Sigma_i=\rho^\Sigma_i/ \rho^\Sigma$ are the interfacial mass fractions, $\bbf^\Sigma= \sum_i y^\Sigma_i \bbf^\Sigma_i$ the total interface body force density and $h^\Sigma$ is the interface enthalpy density. From a kinetic theory, or from a more elaborate continuum theory employing partial momenta (see \cite{9}), the interface driving forces are
\begin{equation}\label{E37}
\dbf^\Sigma_i= \rho^\Sigma_i \na_\Sigma \frac{\mu^\Sigma_i}{T^\Sigma}- \rho^\Sigma_i \frac{\bbf^\Sigma_i -\bbf^\Sigma}{T^\Sigma}+ \frac{y^\Sigma_i}{T^\Sigma} \na_\Sigma \sigma- h^\Sigma_i \na_\Sigma \frac{1}{T^\Sigma}
\end{equation}
with the partial interface enthalpy densities $h^\Sigma_i$. Note that both agree in the isothermal case. Now, equation (\ref{E35}) is used to derive constitutive equations for the $\dbf^\Sigma_i$, after $\dbf^\Sigma_N$, say, is eliminated by means of the constraint
\begin{equation}\label{E38}
\sum\limits^N_{i=1} \dbf^\Sigma_i=0.
\end{equation}
Linear closure yields
\begin{equation}\label{E39}
\dbf^\Sigma_i= - \sum\limits^{N-1}_{k=1} \tau_{ik} (\ubf^\Sigma_k- \ubf^\Sigma_N) \quad\text{ for }\quad i=1, \dots, N-1
\end{equation}
with a positive definite matrix $[\tau_{ik}]$ of interaction coefficients. By arguments fully analogous to those given in \cite{9} for the bulk diffusion case, the assumption of binary interactions, i.e.\ $\tau_{ik}=\tau_{ik}(T^\Sigma, \rho_i^\Sigma, \rho_k^\Sigma) \to 0$ if $\rho^\Sigma_i$ or $\rho^\Sigma_k$ tend to $0$, leads to the Maxwell-Stefan form
\begin{equation}\label{E40}
\dbf^\Sigma_i= - \sum\limits^N_{k=1} f_{ik} \rho^\Sigma_i \rho^\Sigma_k (\ubf^\Sigma_i- \ubf^\Sigma_k) \quad\text{ for }\quad i=1, \dots, N
\end{equation}
with positive, symmetric friction coefficients $f_{ik}$. An equivalent form for molar quantities is
\begin{equation}\label{E41}
-\sum\limits^N_{k=1} \frac{x^\Sigma_k \Jbf^\Sigma_i- x^\Sigma_i \Jbf^\Sigma_k}{D^\Sigma_{ik}}= \dbf^\Sigma_i \quad\text{ for }\quad i=1, \dots, N
\end{equation}
with the interfacial Maxwell-Stefan diffusivities $D^\Sigma_{ik}$. It turns out that inversion of the system (\ref{E41}) together with (\ref{E18}) leads to diffusion fluxes which guarantee positivity of the solutions of the final partial differential equations for the species concentrations; cf.\ \cite{10, 11}. The interface Maxwell-Stefan equations are also derived in \cite{12}, but under the assumption of quasi-stationary hydrodynamics which is not needed here.

The third term in (\ref{E29}) represents the entropy production due to interface chemical reactions, where $R^\Sigma_a$ is the rate of the $\rm{ a^{th}}$ reaction and $\mathcal{A}^\Sigma_a$ the associated affinity. For chemical reactions, a linear closure is usually not appropriate. Since interface chemistry is not in the focus of the present work, we refer to \cite{9} for a non-linear closure of chemical reaction rates which, in simplest cases, resembles the well-known mass-action kinetics.

The $\rm{4^{th}}$ and $\rm{5^{th}}$ binary products in (\ref{E29}) correspond to the dissipation due to momentum transfer between the bulk phases and the interface. Linear closure yields the Navier slip conditions
\begin{equation}\label{E42}
(\vbf^\pm -\vbf^\Sigma)_{||}+ \alpha^\pm(\Sbf^\pm \cdot \nbf^\pm)_{||}=0 \quad\text{ with }\quad \alpha^\pm \geq 0.
\end{equation}
A common choice is $\alpha^\pm=0$ which leads to continuous tangential velocities at $\Sigma$ as used in (\ref{E4}) in the standard model.

The $\rm{6^{th}}$ and $\rm{7^{th}}$ binary products in (\ref{E29}) refer to entropy production during energy transmission from the bulk phases to the interface.
Linear closure yields the relations
\begin{equation}\label{E43}
\frac{1}{T^\Sigma}- \frac{1}{T^\pm}=\beta^\pm \left( \dot{m}(e^\pm+ \frac{p^\pm}{\rho^\pm})+ \qbf^\pm \cdot \nbf^\pm \right) \quad\text{ with }\quad \beta^\pm \geq 0.
\end{equation}
A common choice is $\beta^\pm=0$ which leads to continuous temperatures at $\Sigma$.

The final two binary products in (\ref{E29}) describe the entropy production due to transfer of partial mass across the interface. If cross-effects between the species are ignored, linear closure gives
\begin{equation}\label{E44}
\dot{m}^{\pm,\Sigma}_i= \gamma_i^\pm \left(\frac{\mu^\pm_i}{T^\pm}- \frac{\mu^\Sigma_i}{T^\Sigma}+ \frac{1}{T^\Sigma} \left( \frac{(\vbf^\pm-\vbf^\Sigma)^2}{2}-\nbf^\pm \cdot \frac{\Sbf^{\pm,\rm visc}}{\rho^\pm}\cdot \nbf^\pm \right)\right)
\end{equation}
with $\gamma_i^\pm \geq 0$.
Equation (\ref{E44}) shows that mass transfer is driven by chemical potential differences but, in addition, a kinetic term and viscous forces contribute to the driving force. In the limiting cases, as $\gamma_i^\pm \to\infty$, one obtains
\begin{equation}\label{E45}
\frac{\mu^\pm_i}{T^\pm}= \frac{\mu^\Sigma_i}{T^\Sigma}- \frac{1}{T^\Sigma} \left(\frac{(\vbf^\pm-\vbf^\Sigma)^2}{2}-\nbf^\pm \cdot \frac{\Sbf^{\pm,\rm visc}}{\rho^\pm}\cdot \nbf^\pm \right).
\end{equation}
These limiting cases are referred to as those of vanishing (one-sided) interfacial resistance against mass transfer. If the temperature is assumed to be continuous, often an appropriate assumption, and if the kinetic and viscous terms can be neglected compared to the chemical potentials, then (\ref{E45}) yields
\begin{equation}\label{E46}
\mu^+_i=\mu^-_i=\mu^\Sigma_i;
\end{equation}
cf.\ \cite{1} for an estimation of the strength of the different contributions in (\ref{E45}). Let us note that the first equation in (\ref{E46}), which is usually used to describe the concentration jump at the interface, has to be evaluated with care, since not only the respective one-sided limits of bulk compositions enter the chemical potentials, but also the different pressures:
\begin{equation}\label{E47}
\mu^+_i(T, p^+, x^+_1, \dots, x^+_N)= \mu^-_i(T, p^-, x^-_1, \dots, x^-_N).
\end{equation}
Since the pressure also has a jump at $\Sigma$ and the height of this jump depends on the curvature, the pressure dependence of the chemical potentials introduces a curvature influence into (\ref{E47}), implying in particular that smaller bubbles display an increased solubility. In the context of thermally driven phase transfer, the same effect is described by the Kelvin equation.

Note that if (\ref{E45}) or (\ref{E47}) is employed instead of (\ref{E44}), say, than the mass transfer rates are no longer explicitly determined from the driving differences in chemical potentials, but only follow indirectly from the bulk concentration profiles at the interface for the diffusive contribution and from the total mass transfer rate for the convective part. If $\dot{m}$ is also not known explicitly, which is usually the case since continuity of the temperatures is assumed, then the convective part can still be accounted for, but follows from the solution of a linear system of equations; cf.\ \cite{2,13}.

In order to understand the influence of surface coverage by surfactant to the mass transfer of another chemical species, we have to stick to a general closure like (\ref{E44}). But we prefer a non-linear closure in analogy to chemical reactions, viewing the mass transfer from the bulk phases to the interface as ad- and desorption processes, since the system can be far away from mass transfer equilibrium. For technical simplicity we neglect the kinetic and viscous terms, i.e.\ we exploit the reduced products
\begin{equation}\label{E48}
\zeta^\pm_{\rm TRANS}= \frac{1}{T} \sum\limits_{i=1}^N \big(s^{ad,\pm}_i- s^{de,\pm}_i\big) \left(\mu^\pm_i- \mu^\Sigma_i\right)
\end{equation}
to derive the closure relations
\begin{equation}\label{E49}
\ln \frac{s^{ad,\pm}_i}{s^{de,\pm}_i}= \frac{a^\pm_i}{RT} \big(\mu^\pm_i-\mu^\Sigma_i \big) \quad\text{ with }\quad a^\pm_i \geq 0.
\end{equation}
Note that one of the rates, either the ad- or the desorption rate, has to be modeled based on a micro-theory or experimental knowledge. Then the other rate follows from (\ref{E49}). Below, we will let $a^\pm_i=1$ for simplicity which already suffices to obtain interesting new results. We first consider the case of a soluble surfactant which, for simplicity, is only present in $\Omega^+$, say, and on $\Sigma$. Desorption is often more easy to model, where the simplest rate function is
\begin{equation}\label{E50}
s^{de}_i= k^{de}_i x^\Sigma_i
\end{equation}
with the interfacial molar fraction $x^\Sigma_i=c^\Sigma_i/c^\Sigma$. According to (\ref{E49}) with $a^\pm_i=1$, the associated adsorption rate is
\begin{equation}\label{E51}
s^{ad}_i= k^{de}_i x^\Sigma_i \exp \big(\frac{\mu^+_i- \mu^\Sigma_i}{RT} \big).
\end{equation}
To achieve a concrete result, assume ideal mixtures both in the bulk and on the interface, i.e.\ \begin{equation}\label{E52}
\mu^\pm_i(T,p,x_1, \dots, x_{N-1})= g^\pm_i(T,p)+ RT \ln x^\pm_i
\end{equation}
with $g^\pm_i(T,p)$ denoting the bulk Gibbs free energy of component $A_i$ under the temperature and pressure of the mixture and
\begin{equation}\label{E53}
\mu^\Sigma_i(T,p^\Sigma,x^\Sigma_i, \dots, x^\Sigma_{N-1})= g^\Sigma_i(T,p^\Sigma)+ RT \ln x^\Sigma_i
\end{equation}
with $g^\Sigma_i(T,p^\Sigma)$ denoting the surface Gibbs free energy of component $A_i$ under the temperature and surface tension of the mixture. Insertion of the chemical potentials into (\ref{E51}) yields
\begin{equation}\label{E54}
s^{ad}_i= k^{de}_i \exp \big(\frac{g^+_i-g^\Sigma_i}{RT} \big) x^+_i=: k^{ad}_i x^+_i,
\end{equation}
where $k^{ad}_i$ depends in particular on the surface tension. Together, this yields the simplest ad- and desorption rates, leading to the so-called Henry isotherm.

Next, we consider a transfer component $A_i$, like a dissolving gas, which does not accumulate at the interface as a surfactant does, but nevertheless needs to pass through the transmission layer at the phase boundary which is mathematically represented by the sharp interface. In this case, the interface concentration $A_i$ is assumed so small that (\ref{E13}) can be approximated by
\begin{equation}\label{E55}
[\![\dot{m}_i]\!]=0 \quad \Leftrightarrow \quad \dot{m}^{+,\Sigma}_i+ \dot{m}^{-,\Sigma}_i=0.
\end{equation}
Consequently, employing (\ref{E16}) and (\ref{E49}) with $a^\pm_i=1$, we obtain
\begin{equation}\label{E56}
s^{de,+}_i \left(\exp \big( \frac{\mu^+_i-\mu^\Sigma_i}{RT}\big)-1\right)+ s^{de,-}_i \left(\exp \big( \frac{\mu^-_i-\mu^\Sigma_i}{RT}\big)-1\right)=0.
\end{equation}
Employing the same simplifying assumption of ideal mixtures and (\ref{E50}), this implies
\begin{equation}\label{E57}
k^{de,+}_i \left(\exp \big( \frac{g^+_i-g^\Sigma_i}{RT}\big)x^+_i-x^\Sigma_i\right)+ k^{de,-}_i \left(\exp \big( \frac{g^-_i-g^\Sigma_i}{RT}\big)x^-_i-x^\Sigma_i\right)=0,
\end{equation}
hence
\begin{equation}\label{E58}
\displaystyle
x^\Sigma_i= \frac{k^{de,+}_i \exp \big(\frac{g^+_i-g^\Sigma_i}{RT}\big)x^+_i
+ k^{de,-}_i \exp \big( \frac{g^-_i-g^\Sigma_i}{RT}\big)x^-_i}{k^{de,+}_i+ k^{de,-}_i}.
\end{equation}
Inserting this value into the first expression in (\ref{E57}), equation (\ref{E16}) yields
\begin{equation}\label{E59}
\dot{m}^{+,\Sigma}_i= \frac{k^{de,+}_ik^{de,-}_i}{k^{de,+}_i+ k^{de,-}_i} \exp \big(- \frac{g^\Sigma_i}{RT}\big) \left( \exp \big( \frac{g^+_i}{RT}\big) x^+_i- \exp \big( \frac{g^-_i}{RT}\big)x^-_i\right).
\end{equation}
Let us compare (\ref{E59}) with a closure which does not account for the interface concentrations. In this case, (\ref{E55}) is directly build into the entropy production such that only a single binary product per species remains, namely
\begin{equation}\label{E60}
\zeta^\Sigma_{\rm TRANS}=- \frac{1}{T} \sum\limits^N_{i=1} \dot{m}_i [\![\mu_i+ \frac{(\vbf-\vbf^\Sigma)^2}{2}- \nbf_\Sigma \cdot \frac{\Sbf^{\rm visc}}{\rho}\cdot \nbf_\Sigma]\!].
\end{equation}
Neglecting again the kinetic and viscous terms, the non-linear closure analogous to (\ref{E49}) yields
\begin{equation}\label{E61}
\dot{m}^{+,\Sigma}_i(=- \dot{m}^{-,\Sigma}_i)= k_i \exp \big( -\frac{g^-_i}{RT} \big) \left(\exp \big(\frac{g^+_i}{RT}\big) x^+_i- \exp\big(\frac{g^-_i}{RT}\big)x^-_i\right),
\end{equation}
where the ideal mixture assumption is used.
The most important difference to (\ref{E59}) is that there, the mass transfer rate is influenced by the surface tension via the surface Gibbs free energy. In contrast to (\ref{E61}), this allows to account for the effect of surfactants on the mass transfer of the considered transfer component.
For example, assume the surface equation of state to be
given as
\begin{equation}\label{surfpress}
p^\Sigma = RT \sum_{i=1}^{N-1} c^\Sigma_i \, + \, K^\Sigma \Big(  \frac{c^\Sigma_N}{c^\Sigma_{\rm ref}} -1 \Big).
\end{equation}
The basis for this equation of state is that the interface with its surface tension in the clean state is built by component $A_N$ (the solvent, say) as the phase boun\-dary between a liquid and a vapor phase.
The phase boundary is modeled as a compressible interface phase with compressibility $K^\Sigma$.
Additional components $A_1,\ldots ,A_{N-1}$ are (potentially) present in both bulk phases as well
as on the interface.
Then the surface analog of the construction of a consistent free energy as explained in
\S 15 in \cite{9} yields the corresponding interface free energy as
\begin{equation}
\rho^\Sigma \psi^\Sigma =  K^\Sigma +
\Big( K^\Sigma \frac{c^\Sigma_N}{c^\Sigma_{\rm ref}} + RT \!\sum_{i=1}^{N-1}\! c^\Sigma_i \Big)
\! \Big(\! \ln \! \big( \frac{c^\Sigma_N}{c^\Sigma_{\rm ref}} + \frac{RT}{K^\Sigma} \!\sum_{i=1}^{N-1}\! c^\Sigma_i \big) -1 \Big)
+ RT \sum_{i=1}^N c_i^\Sigma \ln x_i^\Sigma \!.
\end{equation}
The corresponding (molar based) chemical potentials then are
\begin{equation}
\mu_i^\Sigma =  R T  \ln \big( 1 + \frac{p^\Sigma}{K^\Sigma} \big) + RT \ln x_i^\Sigma
\quad \mbox{ for } i=1,\ldots , N-1
\end{equation}
and
\begin{equation}
\mu_N^\Sigma =  \frac{K^\Sigma}{c^\Sigma_{\rm ref}}  \ln \big( 1 + \frac{p^\Sigma}{K^\Sigma} \big) + RT \ln x_N^\Sigma.
\end{equation}
Hence the ideal mixture form according to (\ref{E53}) results with
\begin{equation}\label{E62}
g^\Sigma_i(T,p^\Sigma)= RT \ln \big( 1 + \frac{p^\Sigma}{K^\Sigma} \big) \quad \mbox{ for } i=1,\ldots , N-1.
\end{equation}
Insertion of (\ref{E62}) into (\ref{E59}) implies the relation
\begin{equation}
\dot{m}^{+,\Sigma}_i= \frac{k^{de,+}_ik^{de,-}_i}{k^{de,+}_i+ k^{de,-}_i}
\frac{1}{1 + p^\Sigma /K^\Sigma}
\left( \exp \big( \frac{g^+_i}{RT}\big) x^+_i- \exp \big( \frac{g^-_i}{RT}\big)x^-_i\right).
\end{equation}
Taking the clean surface as the reference state, this yields
\begin{equation}\label{E65}
\dot{m}^{\rm contam}_i=\frac{1+ p^\Sigma_{\rm clean}/K^\Sigma}{1+ p^\Sigma_{\rm contam}/K^\Sigma} \, \dot{m}^{\rm clean}_i
=\frac{K^\Sigma - \sigma_{\rm clean}}{K^\Sigma - \sigma_{\rm contam}} \, \dot{m}^{\rm clean}_i
\end{equation}
for the mass transfer in a system contaminated by surfactant, given as a relation to that in the clean system.
The specific relation (\ref{E65}) evidently results from very strong
assumptions which are not realistic in particular for high surfactant concentrations.
If the interface (or parts of it) is covered by a densely packed surfactant layer it becomes (locally) almost incompressible.
In this case, if the interfacial mass density is assumed to be constant, the surface pressure dependence of the surface
chemical potentials becomes linear. Then, if the interface mixture is still assumed to be ideal, the
(molar based) chemical potentials are of the form
\begin{equation}\label{E65b}
\mu_i^\Sigma (T , p^\Sigma, x_k^\Sigma) = f_i^\Sigma (T )
+ M_i \frac{p^\Sigma}{\rho^\Sigma} + RT \ln x_i^\Sigma.
\end{equation}
Hence (\ref{E53}) again holds but with
\begin{equation}\label{E62b}
g^\Sigma_i(T,p^\Sigma)= f_i^\Sigma (T ) + M_i \frac{p^\Sigma}{\rho^\Sigma}.
\end{equation}
Via (\ref{E59}), this corresponds to a reduction of the mass transfer rate by an exponential
damping factor of Boltzmann type, i.e.\ a factor
of the form $k \exp (-a \, p^\Sigma /RT)$, in accordance with the energy barrier model due to Langmuir; see \cite{Langmuir}, \cite{Ciani}
and the references given there. Experimental data for the transfer of $\rm{CO_2}$ from Taylor bubbles under the influence of different surfactants in \cite{14} also support the fact that the surface pressure of the contaminated system - not, in the first place, the surfactant concentration - determines the mass transfer reduction.

The description above provides a means to include mass transfer hindrance induced by surface coverage into detailed numerical simulations: if the interfacial resistance of the clean system against mass transfer is negligible, say, one can compute the local mass transfer rates from the bulk concentration gradients at the interface, accounting for Henry's law to model the concentration jumps, and then correct the transfer rates via (\ref{E65}) or a more elaborate modeling using (\ref{E59}) and appropriate interface free energies. Notice that a second significant effect of surface coverage is the Marangoni-convection resulting from the non-homogeneous surface tension in (\ref{E22}) which often leads to an immobilization of the interface at the rear end of moving bubbles. For spherical bubbles, the latter is typically accounted for by the stagnant cap model; for this as well as for a more detailed but still simplified model of the immobilization effect see \cite{15} and the references given there. In future work we will further elaborate on the above approach to describe mass transfer under the influence of adsorbed surfactant, considering more involved models of the interfacial free energy, also taking into account electrical effects due to, say, ionic surfactant or salts.
\section*{The Vicinity of the Interface}
The immediate consequence of the processes at the interface is the occurrence of possibly strong gradients in the adjacent bulk regions. This concerns both phases and can lead to local depletion of a certain species due to fast transfer through the interface into the other bulk phase. In this context it should be mentioned that mass transfer problems for multicomponent systems are always conjugate problems. This is nicely demonstrated by the experiments on CO$_2$-transfer from Taylor bubbles into water in \cite{16} in which case a pure CO$_2$-bubble is present initially, but during its dissolution into the water, other gas components dissolved in the water phase lead to mass transfer into the bubble such that it does not completely disappear but a gas bubble, free of CO$_2$, finally persists.

The dissolution of a chemical component in the bulk phase into which it is transferred typically leads to a release of heat (heat of mixing) which can induce locally non-isothermal conditions. This effect is of course stronger for smaller bubbles since the specific area scales as 1/$R$ for a spherical bubble of radius $R$. This may explain that small CO$_2$-bubbles with diameter below 100 mm show faster rise velocities than predicted by the classical Hadamard-Rybczinski solution although the latter assumes a fully mobile interface; see \cite{17}, where a local reduction of the liquid viscosity due to a locally increased temperature is mentioned as a possible explanation. This effect will be even much stronger in case of reactive mass transfer with fast exothermal chemical reaction, since in this case the chemical reaction will occur mainly in the region near the interface.

The presence of fast chemical reactions will also amplify another well-known phenomenon which is a severe challenge for so-called direct numerical simulations, i.e.\ the approximate numerical solution of the mathematical model with such a fine resolution in time and space that all relevant time and length scales are sufficiently resolved for an accurate and grid independent solution. This phenomenon is the occurrence of extremely thin concentration boundary layers due to advection-dominated transport. To explain this in some more detail, consider the dimensionless form of the bulk transport equation for molar mass. Here it suffices to consider the simplest form from the standard model, i.e.\ \begin{equation}
\partial_t^\ast c_i^\ast + \nabla^\ast \cdot (c_i^\ast \vbf^\ast ) =
\frac{1}{{\rm Re} \, {\rm Sc}_i} \Delta^\ast c_i^\ast,
\end{equation}
where chemical reactions are not included for simplicity. The dimensionless factor in front of the diffusion term is much smaller than 1 in most technically relevant cases, showing that advective transport is much stronger than diffusive one. But even more severe, the factor in front of the Laplacian is much smaller then 1/Re, which governs the diffusion of momentum. Recall that typical Schmidt numbers for dissolved components in liquids are of the order of 1000 or even a few magnitudes larger for macro-molecular components, e.g. proteins or large surfactant molecules. The effect of the Schmidt number is the occurrence of finest structures in the concentration fields, like local lamellar sheets or boundary layers, with typical length scale $l_{\rm diff}$ down to the Batchelor length scale given as
\begin{equation}
l_{\rm diff} = l_{\rm conv} / \sqrt{{\rm Sc}_i},
\end{equation}
where $l_{\rm conv}$ denotes the smallest length scale of the velocity field. This was derived in \cite{18} for scalar transport in turbulent flows, in which case $l_{\rm conv}$ is the Kolmogorov length, but is also valid as a lower bound for the length of the smallest persisting structures in a scalar distribution as shown computationally in \cite{19} for mixing in a T-shaped micromixer. At present, direct numerical simulations of mass transfer at gas bubbles, say, in fully three-dimensional cases even for moderate Reynolds numbers are not feasible for realistic Schmidt numbers. This is true even more for the case of reactive mass transfer. The solution to this ``High Schmidt Number Problem (HSNP)'' is one of the largest numerical challenges in this field.
\section*{The Gas Phase}
In virtually every two-phase fluid system, the gas phase will be composed of more than one chemical component; even if a pure gas bubble is placed in a degassed liquid, the gas phase will receive liquid vapor during the bubble evolution. Often three or more constituents will be present in the gas phase in which case diffusion cross-effects can become relevant. This is a well-known fact which leads to phenomena like up-hill diffusion as shown in the classical experiment by Duncan and Toor \cite{20}. A proper description of multicomponent diffusion with cross-effects and non-ideality effects is possible via the Maxwell-Stefan equations together with a proper modeling of the chemical potentials including especially their pressure dependence. The bulk version of the Maxwell-Stefan equations are completely analogous to (\ref{E18}), (\ref{E37}) and (\ref{E41}) above, only that $\sigma$ is to be replaced by $-p$ in (\ref{E37}). For more details on the Maxwell-Stefan description of multicomponent diffusion see \cite{9, 21, 22}.
A second important weak point of the standard model is the assumption of isochoric flow inside the gas phase, i.e.\ zero divergence of the barycentric velocity field. This excludes volume variations of the bubble as a reaction to changes in pressure, but the latter can be significant in particular for bubbles rising due to buoyancy, since such bubbles can experience strong variations of hydrostatic pressure which typically leads to expansion of the bubbles during their rise, accompanied by a change of the constituents' chemical concentrations. To include such pressure effects into the modeling, one should note that incompressibility of the gas phase is usually introduced via the low Mach number (Ma) approximation. The latter employs an expansion of the balanced quantities as power series in Ma and, by equating coefficients for corresponding powers, the pressure contributions $p_0$ and $p_1$ are ruled out, while $p_2$ is the pressure which remains as a Lagrange parameter in (\ref{E2}), taking care of the zero divergence condition. But a closer look at the low Mach number approximation shows that $p_0$ and $p_1$ are not constant: They have zero gradients, but are still allowed to depend on time. This indicates a possible way to include the pressure dependence inside the gas phase.
\section*{The Liquid Phase}
As in the gas phase, the assumption of incompressibility in the standard model can also be a weak point for the liquid phase. First, incompressibility is often even imposed via the assumption of constant mass density. At this point, note that the physical definition of incompressibility for a single component material states that the mass density given as a function of temperature and pressure (which then is the appropriate form of the thermal equation of state), shows no dependence on pressure. This can be adapted to multicomponent mixtures, but such extensions will neither lead to constant total mass density nor to divergence free barycentric velocity fields; cf.\ \cite{23}. Concerning the mass density variations, a Boussinesq-type approximation seems possible but a rigorous and thermodynamically consistent modeling of such composition-induced volume effects is currently lacking. Such volume effects in the liquid case will be more severe in case of reactive mass transfer since the chemical reactions transform certain educts with their specific volume fraction into products which usually have different specific partial volumes.

If chemical reactions take place in the liquid, for instance in bubble column processes, large sets of constituents are commonly involved. Despite the volume effects mentioned above, this can lead to another significant complication occurring especially in aqueous solutions but also in other polarizable liquids: While the transfer components are uncharged, ionic species usually get involved during the course of chemical transformation. This couples the diffusive mass fluxed due to the appearance of Coulomb forces exerted by the electric field which is generated by the ions themselves. To account for these effects requires a coupling of the Maxwell-Stefan approach, say, to at least a quasi-static version of the Maxwell equations from electromagnetics, i.e.\ the Poisson equation to model the quasi-stationary electrical field generated by the ions. Since a detailed discussion of these model extensions is out of the scope of the present paper, we only refer to the recent publications \cite{24, 25, 26}, but note that electrical effects can also play a role directly at fluid interfaces. For the latter, see also the recent Ph.D.\ thesis \cite{30}.
\section*{The Outer Boundaries}
A sound model requires sensible conditions at the outer boundaries of the balanced region, which also includes a sound choice of appropriate outer boundaries itself. If possible, it is favorable  to choose the outer boundaries to coincide with physical phase boundaries, like impermeable walls or a free liquid surface. In this case knowledge of the physics can be exploited to formulate reasonable boundary conditions in pretty much the same way as it has been done above for the fluid interface, i.e.\ by a combination of interfacial mass, momentum and energy balances with closure relations being motivated by the local entropy production rate. A severe complication for the mathematical description is the fact that the position of such physical outer boundaries is usually not fixed but will (slightly, say) vary due to the acting forces. In other words, the boundaries are usually pressure controlled, while almost all detailed mathematical descriptions employ volume control by fixing the spatial domain in which the considered processes take place.

In detailed or even direct numerical simulations one often cannot extend the domain up to physical outer boundaries, but has to introduce artificial boundaries somewhere inside the bulk phases. At such non-physical boundaries the formulation of appropriate boundary conditions, so-called Artificial Boundary Conditions (ABCs), is a field in its own and an important topic in Computational Fluid Dynamics. Let us note in passing that the use of perfect slip, i.e.\ \begin{equation}
\vbf \cdot \nbf =0 \quad \mbox{ and } \quad ({\bf S} \nbf )_{||} =0,
\end{equation}
models the boundary as impermeable which is in general wrong at an artificial inner surface inside a bulk phase. While quite some research has been devoted to the modeling of ABCs at outflow boundaries for single-phase flows (see, e.g., \cite{27, 28}, the references given there and in \cite{29}), much less is known for ABCs in two-phase flow situations. For rising or settling single fluid particles, an artificial boundary condition for lateral domain boundaries was derived in \cite{29} which allows flow trajectories to exit and re-enter the computational domain. This allows for a reduction of the computational domain by about a factor of two in each lateral direction, thereby enabling a finer resolution near the interface, where it is crucial because of the HSNP mentioned above.
\section*{Conclusion}
A multitude of physical phenomena can occur at fluid interfaces under strongly dynamical conditions. The potential of direct numerical simulations for gaining deep insights into such processes at fluid interfaces can only be fully exploited if the employed models provide a complete and sound description of the multi-scale multi-physics. For this purpose, the existing continuum mechanical models have to be critically assessed, improved and extended. The entropy principle provides a rational framework for such model extensions and its rigorous exploitation leads to deep insights into, e.g., mass transfer across contaminated fluid interfaces.
\section*{Acknowledgment}
The author gratefully acknowledges financial support provided by the German Research Foundation (DFG)  within the Priority Program SPP1740 ``Reactive Bubbly Flows''. He also would like to thank
Wolfgang Dreyer (WIAS, Berlin) for many helpful discussions and for his critique on a prior
version of this paper.

\section*{Appendix: Interfacial Entropy Production}
We explain the main steps of a rigorous derivation of the entropy production in multi-component two-phase systems with non-vanishing interface mass densities. The starting point are the entropy balances in the bulk phases and on the interface, which read as
\begin{equation}\label{E69}
\pa_t(\rho s)+ \na \cdot (\rho s \vbf+\Phibf)=\zeta
\end{equation}
in the bulk phases with $s$ the specific entropy, $\Phibf$ the entropy flux and $\zeta$ the bulk entropy production, and
\begin{equation}\label{E70}
\pa^\Sigma_t(\rho^\Sigma s^\Sigma) + \na_\Sigma \cdot (\rho^\Sigma s^\Sigma \vbf^\Sigma+ \Phibf^\Sigma)+ [\![\rho s(\vbf-\vbf^\Sigma)+\Phibf]\!] \cdot \nbf_\Sigma=\zeta^\Sigma
\end{equation}
with the specific interface entropy $s^\Sigma$, the interfacial entropy flux $\Phibf^\Sigma$ and the interfacial entropy production $\zeta^\Sigma$. The second law of thermodynamics now states that
\begin{equation}\label{E71}
\zeta^\pm \geq 0 \quad\text{ and }\quad \zeta^\Sigma \geq 0 \quad\text{ for any thermodynamic process}.
\end{equation}
The entropy densities are objective scalars which have to be modeled as material dependent quantities, where we consider the simplest set of primitive variables:
\begin{equation}\label{E72}
\rho s=\rho s(\rho e, \rho_1, \dots, \rho_N) \quad\text{ and }\quad \rho^\Sigma s^\Sigma=\rho^\Sigma s^\Sigma(\rho^\Sigma e^\Sigma, \rho^\Sigma_1, \dots, \rho^\Sigma_N).
\end{equation}
The partial entropy derivatives have a clear physical meaning, which leads to the following definitions:
\begin{equation}\label{E73}
\frac{1}{T}:= \frac{\pa \rho s}{\pa \rho e}, \quad - \frac{\mu_i}{T}:= \frac{\pa \rho s}{\pa \rho_i} \quad\text{and}\quad \frac{1}{T^\Sigma}:= \frac{\pa \rho^\Sigma s^\Sigma}{\pa \rho^\Sigma e^\Sigma}, \quad - \frac{\mu^\Sigma_i}{T^\Sigma}:= \frac{\pa \rho^\Sigma s^\Sigma}{\pa \rho^\Sigma_i}.
\end{equation}
Using the chain rule, the balance equations (\ref{E69}) and (\ref{E70}) are used to express the entropy productions by means of the respective right-hand sides, where all time derivatives are eliminated by means of the corresponding balance equations. For this purpose the internal energy balance is required which reads
\begin{equation}\label{E74}
\pa_t(\rho e)+ \na \cdot (\rho e \vbf +\qbf)= \na\vbf:\Sbf+ \sum\limits^N_{i=1} \jbf_i \cdot \bbf_i
\end{equation}
in the bulk phases with $e$ the specific energy, $\qbf$ the heat flux and $\bbf_i$ the individual body forces acting on the species $A_i$, and
\begin{equation}\label{E75}
\begin{split}
&\pa^\Sigma_t (\rho^\Sigma e^\Sigma)+\na_\Sigma \cdot(\rho^\Sigma e^\Sigma \vbf^\Sigma+ \qbf^\Sigma)
 + [\![\dot{m}h+ \frac{(\vbf-\vbf^\Sigma)^2}{2}-\nbf_\Sigma \cdot \frac{\Sbf^{\rm visc}}{\rho}\cdot \nbf_\Sigma]\!]\\
& -[\![(\vbf-\vbf^\Sigma)_{||} \cdot(\Sbf^{\rm visc} \nbf_\Sigma)_{||}]\!]+ [\![\qbf \cdot \nbf_\Sigma]\!]= \na_\Sigma \vbf:\Sbf^\Sigma + \sum\limits^N_{i=1} \jbf^\Sigma_i \cdot \bbf^\Sigma_i
\end{split}
\end{equation}
on the interface with the specific bulk enthalpy $h=e+ \frac{p}{\rho}$,
the interface internal energy density $\rho^\Sigma e^\Sigma$, the interfacial heat flux $\qbf^\Sigma$ and the individual body forces $\bbf^\Sigma_i$. After lengthy but straightforward calculations, this yields
\begin{equation}\label{E76}
\begin{split}
\zeta= & \na \cdot(\Phibf- \frac{\qbf}{T}+ \sum\limits^N_{i=1} \frac{\mu_i \jbf_i}{T})- \frac{1}{T} (\rho e + p-\rho sT- \sum\limits_{i=1}^N \rho_i \mu_i) \na \cdot \vbf\\
& + \qbf \cdot \na \frac{1}{T}+ \frac{1}{T} \na \vbf:\Sbf^{\rm visc}- \sum\limits_{i=1}^N \jbf_i \cdot (\na \frac{\mu_i}{T}- \frac{\bbf_i}{T})- \frac{1}{T} \sum\limits_{a=1}^{N_R} R_a \mathcal{A}_a
\end{split}
\end{equation}
for the bulk entropy production, where $p$ is the thermodynamic pressure, $R_a$ are the molar reaction rates of the ${\rm a^{th}}$ chemical bulk reaction and $\mathcal{A}_a$ the associated affinities defined as
\begin{equation}\label{E77}
\mathcal{A}_a = \sum\limits^N_{i=1} M_i \mu_i \nu^a_i
\end{equation}
with the stoichiometric coefficients $\nu^a_i$; see \cite{9} for further details. For the interfacial entropy production we obtain
\begin{align}
\label{E78}
	& \zeta^\Sigma =\na_\Sigma \cdot(\Phibf^\Sigma- \frac{\qbf^\Sigma}{T^\Sigma}+ \sum\limits^N_{i=1} \frac{\mu^\Sigma_i \jbf^\Sigma_i}{T^\Sigma})\\
 & - \frac{1}{T^\Sigma}(\rho^\Sigma e^\Sigma -\sigma -\rho^\Sigma s^\Sigma T^\Sigma - \sum\limits^N_{i=1} \rho^\Sigma_i \mu^\Sigma_i) \na_\Sigma \cdot \vbf^\Sigma
 + \qbf^\Sigma \cdot \na_\Sigma \frac{1}{T^\Sigma}\nonumber\\
	&+ \frac{1}{T^\Sigma} \na \vbf^\Sigma : \Sbf^{\Sigma,\rm visc}- \sum\limits^N_{i=1} \jbf^\Sigma_i \cdot \left( \na_\Sigma \frac{\mu^\Sigma_i}{T^\Sigma}- \frac{\bbf^\Sigma_i}{T^\Sigma} \right)- \sum\limits^{N_R}_{a=1} R^\Sigma_a \mathcal{A}^\Sigma_a\nonumber\\
	&- \frac{1}{T^\Sigma} (\vbf^+-\vbf^\Sigma)_{||} \cdot (\Sbf^+ \nbf^+)_{||}- \frac{1}{T^\Sigma} (\vbf^- -\vbf^\Sigma)_{||} \cdot (\Sbf^- \nbf^-)_{||}\nonumber\\
	&+ \left( \frac{1}{T^\Sigma}- \frac{1}{T^+}\right) \left(\dot{m}(e^+ + \frac{p^+}{\rho^+})+ \qbf^+ \cdot \nbf^+ \right)\nonumber\\
 & + \left( \frac{1}{T^\Sigma}- \frac{1}{T^-}\right) \left( \dot{m}(e^-+ \frac{p^-}{\rho^-})+ \qbf^- \cdot \nbf^-\right)\nonumber\\
	&+ \sum\limits^N_{i=1} \dot{m}^{+,\Sigma}_i \left(\frac{\mu^+_i}{T^+}- \frac{\mu^\Sigma_i}{T^\Sigma}+ \frac{1}{T^\Sigma} \Big(\frac{(\vbf^+-\vbf^\Sigma)^2}{2}-\nbf^+ \cdot \frac{\Sbf^{+,\rm visc}}{\rho^+} \cdot \nbf^+\Big)\right)\nonumber\\
	&+ \sum\limits^N_{i=1} \dot{m}^{-,\Sigma}_i \left(\frac{\mu^-_i}{T^-}- \frac{\mu^\Sigma_i}{T^\Sigma}+ \frac{1}{T^\Sigma} \Big(\frac{(\vbf^--\vbf^\Sigma)^2}{2}-\nbf^- \cdot \frac{\Sbf^{-,\rm visc}}{\rho^-} \cdot \nbf^-\Big)\right).\nonumber
\end{align}
The entropy fluxes are constitutive quantities which have to be modeled as objective vectors in such a way that the resulting entropy productions become sums of binary products, one for each dissipative mechanism. This leads to the constitutive relations
\begin{equation}\label{E79}
\Phibf= \frac{\qbf}{T}- \sum\limits^N_{i=1} \frac{\mu_i \jbf_i}{T} \quad\text{ and }\quad \Phibf^\Sigma= \frac{\qbf^\Sigma}{T^\Sigma}- \sum\limits^N_{i=1} \frac{\mu^\Sigma_i \jbf^\Sigma_i}{T^\Sigma}
\end{equation}
for the entropy fluxes. Next, in order to satisfy the second law of thermodynamics, the term in front of the (surface) divergence of the (interface) velocity needs to vanish. This leads to the Gibbs-Duhem relations in the bulk and on the interface, i.e.\ \begin{equation}\label{E80}
\rho e+p- \rho s T= \sum\limits^N_{i=1} \rho_i\mu_i \quad\text{ and }\quad \rho^\Sigma e^\Sigma-\sigma-\rho^\Sigma s^\Sigma T^\Sigma= \sum\limits^N_{i=1} \rho^\Sigma_i \mu^\Sigma_i,
\end{equation}
also called bulk, respectively interface Euler relations. Hence the reduced bulk entropy production
\begin{equation}\label{E81}
\zeta= \qbf \cdot \na \frac{1}{T}+ \frac{1}{T} \na \vbf:\Sbf^{\rm visc}- \sum\limits^N_{i=1} \jbf_i \cdot (\na \frac{\mu_i}{T}- \frac{\bbf_i}{T})- \frac{1}{T} \sum\limits^{N_R}_{a=1} R_a \mathcal{A}_a
\end{equation}
and, after specializing to vanishing interface viscosities, the reduced interface entropy production (\ref{E29}) result; see \cite{9} for many more details on the entropy principle as the core part of continuum thermodynamics of chemically reacting fluid mixtures.
For a recent extension to electrochemical interfaces see \cite{30}.
\end{document}